\documentclass[aps,prb,twocolumn,floatfix,showpacs]{revtex4}
\usepackage[dvips]{graphicx}
\usepackage{array}
\usepackage{bm,amssymb,amsmath}

\usepackage{epsf}

\begin{document}


\title{Probable absence of a quadrupolar spin-nematic phase\\
  in the bilinear-biquadratic spin-1 chain}

\author{K.~Buchta, G.~F\'ath, \"O.~Legeza, and J.~S{\'o}lyom}

\affiliation{Research Institute for Solid State Physics and Optics, H-1525
Budapest, P.\ O.\ Box 49, Hungary }

\date{\today}

\begin{abstract}
We study numerically the ground-state phase diagram of the bilinear-biquadratic spin-1
chain near the ferromagnetic instability point, where the existence of a gapped or
gapless nondimerized quantum nematic phase has been suggested. Our results, obtained
by a highly accurate density-matrix renormalization-group (DMRG) calculation are
consistent with the view that the order parameter characterizing the dimer phase
vanishes only at the point where the system becomes ferromagnetic, although the 
existence of a gapped or gapless nondimerized phase in a very narrow parameter
range between the ferromagnetic and the dimerized regimes cannot be ruled out.
\end{abstract}


\maketitle

\section{Introduction}

In the past two decades, a large number of papers were devoted to the study of various
properties of quantum spin chains. This was inspired to a large extent by
Haldane's conjecture \cite{haldane} which states that isotropic antiferromagnetic
spin chains with half-odd-integer or integer spin values behave completely differently.
While the excitations spectrum is gapless in the first case, a gap, the so-called
Haldane gap, is generated in the other case.

The most general form of isotropic coupling for $S=1/2$ spins is the usual
Heisenberg model. For higher spin values, higher powers of the spins may appear
in the Hamiltonian and this gives rise to a richer phase diagram. For $S=1$ spins,
assuming the most general isotropic nearest neighbor interaction, the Hamiltonian
has bilinear and biquadratic terms in the spin operators and it can be written in
the form
\begin{equation}    \begin{split}
{\cal H} & = \sum_i {\cal H}_{i,i+1} \\
    & = \sum_i \big[ \cos\theta ( \bm{S}_i \cdot \bm{S}_{i+1}) +
        \sin\theta ( \bm{S}_i \cdot \bm{S}_{i+1})^2 \big] \,.
\label{eq:ham}
\end{split}   \end{equation}
In this parametrization, $\theta = 0$ corresponds to the antiferromagnetic model
where the spectrum has a finite Haldane gap, while at $\theta = \pm \pi$ the
system is ferromagnetic. In fact both phases have a finite extension in the
parameter space. The massive Haldane phase is stable for $-\pi/4 < \theta <
\pi/4 $, while ferromagnetism exists for $\pi/2<\theta < \pi$ and $-\pi<\theta < -3\pi/4$.
The phase boundaries $\theta = \pm \pi/4$, $-3\pi/4$ and $\pi/2$ are in fact special points,
where the model can be solved exactly.\cite{lai,sutherland,takhtajan,babujan}

The Haldane gap vanishes at $\theta = \pi/4$ and three soft modes appear at
$k=0 $ and $\pm 2\pi/3$. The system remains critical for $\pi/4 < \theta < \pi/2$
 with the soft modes remaining at $k=0,\pm 2\pi/3$.\cite{fath,japan} This
gives rise to a power-law decay of correlations with a $3a$ periodic
oscillation, hence the name ``trimerized'' phase, although the translation
symmetry is not broken.

The gap vanishes also at the other end of the Haldane phase, at $-\pi/4$, but it
reopens for $\theta < -\pi/4$ and a massive dimerized phase with spontaneously broken
translational symmetry is found. A possible definition of the dimer order parameter $D$ is
\begin{equation}
D = \lim_{N\to\infty} |D_{N/2}|; \quad
D_i = \langle{\cal H}_{i-1,i}\rangle - \langle{\cal H}_{i,i+1}\rangle\,,
\label{eq:dimerord}
\end{equation}
where chains are considered with open boundary condition. Dimerization is measured as
the alternation in the bond energy in the middle of a long enough open chain.

The properties of this phase are best known for $\theta = -\pi/2$, where a partial
mapping\cite{parkinson, barber, klumper} to the 9-state quantum Potts model allows
to calculate exactly the ground state energy, the gap, the correlation length, and also the
dimer order.\cite{xian}

\begin{figure}[htb]
\includegraphics[scale=0.35]{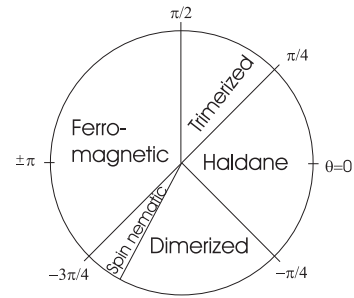}
\vskip .4cm
\caption{Schematic plot of the phase diagram of the bilinear-biquadratic $S=1$
model as a function of $\theta$.}
\label{fig:phase1}
\end{figure}

While the properties and boundaries of the ferromagnetic, Haldane, and critical
``trimerized'' phases have been well established, there has been a long debate
in the literature as for the boundary of the dimerized phase and the eventual
existence of another phase between this dimerized phase and the ferromagnetic
one near $\theta = -3 \pi/4$, as seen in Fig. \ref{fig:phase1}.

By studying fluctuation effects near the end point of the ferromagnetic phase, Chubukov
\cite{chubukov} claimed that there should be a gapped nondimerized nematic phase
between the ferromagnetic and the dimerized phases. In fact, both ferromagnetism and
dimerization involves spontaneous symmetry breaking: for ferromagnetism SU(2) is
broken, whereas for dimerization it is translation invariance. These symmetries are
largely unrelated, and there seems to be no a priori reason  why the two should get
broken hand in hand in one single transition.

\begin{figure}[htb]
\includegraphics[scale=0.35]{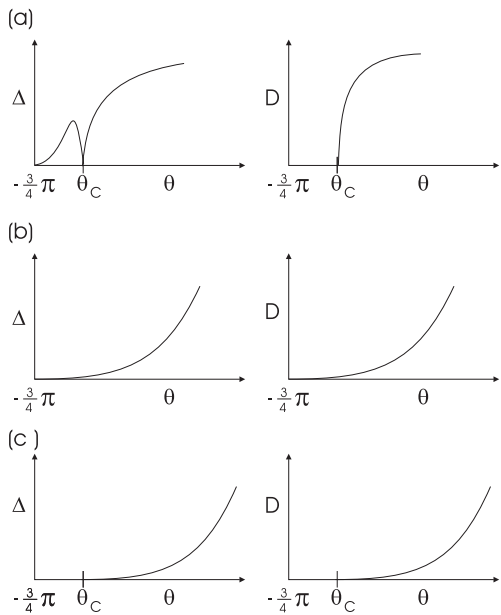}
\vskip .4cm
\caption{Schematic plot of the excitation gap ($\Delta$) and the dimer order
parameter $D$ as a function of $\theta$ around the phase boundary of the
ferromagnetic and dimerized phases. (a) Chubukov's suggestion;
(b) the result of earlier numerical work; (c) the case of a critical nondimerized
phase.}
\label{fig:gaps1}
\end{figure}

According to Chubukov's scenario the dimer order parameter is finite in the dimerized
phase, vanishes together with the gap at a $\theta_{\rm c}$ close to, but definitely above
$-3\pi/4$. The gap reopens for $\theta < \theta_{\rm c}$ and closes again at $-3\pi/4$,
but the dimer order parameter remains zero in this whole range of $\theta$. In this
extended region, the system would have a nondegenerate singlet ground state and 
unbroken translational symmetry. Since the higher dimensional counterpart of Chubukov's 
phase would have quadrupolar order, this phase is usually called the ``spin nematic" phase. 
The behavior of the gap and the dimer order parameter according to this scenario is shown 
schematically in Fig.~\ref{fig:gaps1}(a). Other field theoretic calculations based on a 
nonlinear $\sigma$-model approach for the director field also supported this 
scenario.\cite{ivanov} Note, however, that all these field theoretic
calculations assumed a translation invariant ground state and did not check its stability
against a possible spontaneous breaking of translation invariance.

In our earlier numerical calculations,\cite{fath01,legeza01} where the vanishing of the
gap was studied, we found no evidence for a closing and reopening gap. In these works
both the gap and the dimer order parameter, indicating broken translation invariance,
were reported to vanish at $\theta_{\rm c}=-3\pi/4$ only. This scenario, the absence
of the nematic phase is shown in Fig.~\ref{fig:gaps1}(b).

Recent quantum Monte Carlo \cite{kawashima} and density-matrix renormalization-group 
(DMRG)\cite{lauchli} calculations have indicated that although Chubukov's proposal 
may not be completely correct in a closing and reopening gap, an exotic, \emph{critical} 
phase with quadrupolar correlations may exist between the ferromagnetic and dimerized
phases for $-3\pi/4 < \theta < \theta_{\rm c}$. The behavior of the gap and the dimer
order in this scenario is shown in Fig.\ \ref{fig:gaps1}(c). L\"auchli {\sl et al.}
\cite{lauchli} estimated the value of $\theta_{\rm c}$ to be $ -0.67\pi$.

The aim of the present paper is to find further numerical evidence for the possible
existence of this nondimerized phase. For this purpose we have studied finite
spin chains up to 1000 lattice sites with open (OBC) and periodic boundary
conditions (PBC) for various $\theta$ values using the DMRG method.\cite{white} We
have analyzed the behavior of the excitation gap and the dimer order
parameter given by Eq.~(\ref{eq:dimerord}).

\section{Numerical procedure}

The numerical calculations were performed using the DMRG algorithm. Since
the numerical accuracy is of crucial importance in the present study, this section is
devoted to the problem of how the accuracy of our calculations could be
determined and controlled.

We have performed DMRG calculations both by using the standard
technique,\cite{white} i.e., by keeping the number of block states fixed and
by using the dynamic block state selection (DBSS)
approach.\cite{legeza02,legeza03} All eigenstates of the model have been
targeted independently using two or three DMRG sweeps.

In the standard procedure $M=500-1000$ block states have been used. It was
found that for the largest systems built up of $N = 500-1000$ lattice sites the
truncation error varied in the range $10^{-8}-10^{-9}$ for OBC and
$10^{-5}-10^{-7}$ for PBC. The following numbers are indicative of the accuracy:
for OBC using $M=300$ or $M=500$ block states the ground state energies at $\theta =
- 0.7 \pi$ differ in the fifth digit, $\delta E(300,500)<10^{-5}$, and the
accuracy improved one order of magnitude when $M=1000$ block states
were kept, $\delta E(500,1000) < 10^{-6}$.

The DBSS approach  \cite{legeza02} allows for a more rigorous control of numerical
accuracy, and we set the threshold value of the quantum information loss $\chi$ to
$10^{-8}$. The minimum number of block states $M_{\rm min}$ has been set to $256$.
The entropy sum rule was checked for all finite chain lengths for each DMRG sweeps,
and it was found that the sum rule was satisfied after the second sweep already. The
maximum number of block states varied in the range $600-1400$ for OBC and
$1000-2500$ for PBC, respectively.

After accomplishing the infinite lattice procedure and using White's wave-function
transformation method \cite{stvec} the largest value of the fidelity error of the
starting vector $\Psi_{\rm stv}$, $\delta\epsilon_{\Psi_{\rm stv}} = 1-
\langle \Psi_{\rm T}|\Psi_{\rm stv} \rangle$ was of the order of $10^{-10}$,
where $\Psi_{\rm T}$ is the target state determined by the diagonalization of
the superblock Hamiltonian.

As another test of the accuracy we calculated the dimer order profile $D_i$
using PBC. In theory, this should vanish identically for all finite chain lengths.
Using the DBSS approach with $\chi=10^{-6}$, $M_{\rm min}=256$ for
chains up to $N=200$ sites, in the parameter range $-0.75\pi\le\theta\le-0.5\pi$,
the value obtained for $D_i$ was less than $10^{-5}$ for all $i=1,\dots,N$.
The ratio of the DMRG block energy and the number of bonds within the block
agreed up to 5 digits with the bond energy obtained between the two
DMRG blocks. These results indicate that the finite dimer order parameter
calculated with OBC is probably accurate at least up to 4 digits.

\section{Gap vs dimer order parameter}

In order to obtain the energy gap $\Delta$ and the dimer order parameter $D$ in the
thermodynamic limit $N\rightarrow\infty$, a finite-size scaling analysis has to be
performed. In this section, as a benchmark case, we study in detail the exactly solvable
point $\theta=-\pi/2$. We demonstrate that for OBC, which is usually preferred in DMRG, 
the dimer order parameter is a better quantity to be analyzed, as it provides much more 
accurate results. At the special point $\theta=-\pi/2$ most of the quantities of special 
interest have been determined exactly.\cite{parkinson,klumper,barber,xian} The gap is 
$\Delta_{\rm exact} = 0.173178$, the dimer order parameter reads 
$D_{\rm exact}=1.124378$, and the coherence length is $\xi_{\rm exact}=21.072$.

\subsection{Energy gap}

In a noncritical model with PBC the gap $\Delta(N)$ is expected to scale in leading order 
as
\begin{equation}
    \Delta(N) = \Delta + c{1\over N^{1/2}}\exp(-N/\xi).
    \label{eq:scale_pbc}
\end{equation}
For OBC, however, the corrections are algebraic, and $\Delta(N)$ is expected to vary as
\begin{equation}
    \Delta(N) = \Delta + a/{N^2} + {\cal O}(N^{-4})
    \label{eq:scale_obc}
\end{equation}
where $a$ is a suitable constant. A qualitative argument for this scaling ansatz can be 
given as follows.\cite{sorensen} The magnon dispersion is quadratic around its minimum, 
$\epsilon(k)=\sqrt{\Delta^2+v^2 k^2}$. Due to the boundary condition, however, the 
magnon wavefunction should have nodes on the boundary, which constraints the lowest 
possible magnon momentum to be $k=\pm\pi/N$. Consequently, the lowest possible 
excitation energy (the gap) is $\Delta(N)=\sqrt{\Delta^2+v^2 (\pi/N)^2}\approx \Delta 
+ \pi^2 v^2/2\Delta N^2 +{\cal O}(N^{-4})$, giving Eq.\ (\ref{eq:scale_obc}).

Since in the region of interest the ground state is a singlet and the lowest lying excited
state is in the $S_{\rm tot} = 2$ total spin sector, the excitation gap is calculated from
the energy difference between the lowest lying levels of the $S_{\rm tot}=2$ and
$S_{\rm tot}=0$ sectors
\begin{equation}\label{}
    \Delta(N)=E_{S_{\rm tot}=2}^{(0)}(N)-E_{S_{\rm tot}=0}^{(0)}(N).
\end{equation}

\begin{figure}[htb]
\includegraphics[scale=0.5]{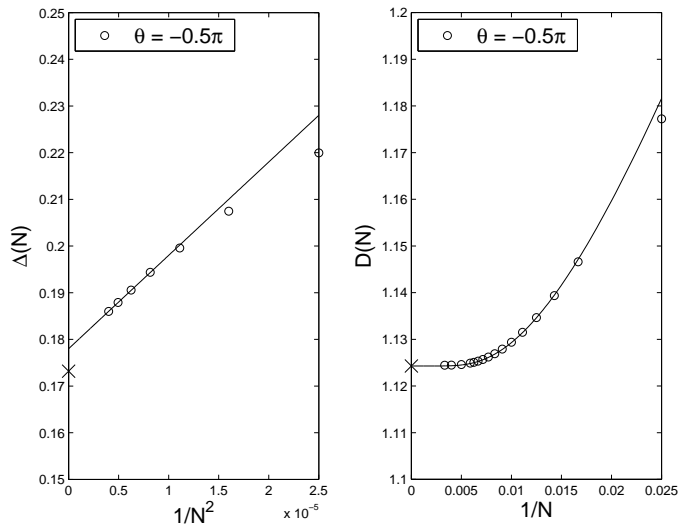}
\caption{The excitation gap ($\Delta$) and dimer order parameter ($D$) at
$\theta=-0.5\pi$ as a function of $1/N^2$. The symbol $\times$ denotes the
exact value. The solid lines are least square fits using the scaling forms discussed in the text.}
\label{fig:dimer_exact}
\end{figure}

We used OBC and the DBSS approach with $\chi=10^{-8}$, $M_{\rm min}=256$.
Our results for the gap as a function of $1/N$ up to $N=500$ are shown in the first panel
of Fig.~\ref{fig:dimer_exact}. A quadratic fit using the form in
Eq.\ (\ref{eq:scale_obc}) for $250\le N\le 500$ yields $\Delta=0.177$ and $a=2300$. 
The gap is about 2\% higher than the exact result.

\subsection{Dimer order}

\begin{figure}[htb]
\includegraphics[scale=0.35]{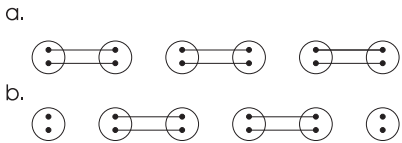}
\caption{Typical valence-bond configurations for chains with even lattice sites.}
\label{fig:valence}
\end{figure}

For finite open chains with an even number of lattice sites the two typical valence-bond
configurations of the dimerized phase are shown in Fig.~\ref{fig:valence}. Due to the
boundary condition these two singlet states are separated by a finite energy gap even in
the thermodynamic limit. The ground state retains a high overlap with the configuration
depicted in Fig.~\ref{fig:valence}(a) as $N\to\infty$. This allows us to measure the
possible dimer order by considering $D(N)$ and take the limit $N\to\infty$. Note that
this is only possible for OPB. For PBC the two possible dimerized configurations mix up
and restore translation invariance for any finite $N$. In this case dimer order could only
be measured by considering the dimer (4-point bond-bond) correlation function. In
fact this was the quantity measured numerically in Ref.\ \onlinecite{lauchli}. Since
calculations with PBC give less accurate results in DMRG than with OPB, we do not
pursue here this approach.

Accordingly, in our numerical calculations we used OBC with $\chi\simeq 10^{-8}~$,
$M_{\rm min}=256$ and determined $D(N)$ as defined in Eq.\ (\ref{eq:dimerord}) at
the center of the chain. Two DMRG sweeps were taken and we checked that the
entropy sum rule was satisfied. Our result is shown in the second panel of
Fig.~\ref{fig:dimer_exact}. The upward curvature of the data points as a function of
$1/N$ is apparent for very short chain lengths ($N\simeq40-80$) and already for
$N>200$ $D(N)$ agrees up to three digits with the infinite chain-length limit.

For noncritical models the characteristic behavior of the system is determined by a finite
correlation length. Therefore, the end effects decay exponentially and the \emph{local}
quantity $D(N)$ is expected to vary in leading order according to
\begin{equation}
    D(N) = D + d N^{-\beta} \exp(-N/2\xi),
    \label{eq:scale_D}
\end{equation}
which is qualitatively similar to the PBC scaling of the gap in Eq.~(\ref{eq:scale_pbc}), 
except that the scaling variable is the distance of the middle of the chain from the boundary, 
$N/2$, and the exponent of the algebraic prefactor is a priori unknown. Nevertheless, 
knowing the exact value of $D$ for $\theta=-0.5\pi$, a least square fit provided $\beta$ 
very close to 1. Using this, our numerical data for $N>60$ can be fitted with 
$D=1.124375$, $\xi=20.2$, and $a=5.9$ (see the solid line in the second panel of 
Fig.~\ref{fig:dimer_exact}). The correlation length is off by 4\%, but the numerical value 
of $D$ has an excellent relative accuracy of $3\times 10^{-6}$.  The exponential 
convergence of local quantities such as the dimer order parameter $D(N)$ makes the 
extrapolation to the thermodynamic limit very reliable. Our general conclusion is 
that---because of their different scaling behavior---the dimer order parameter is a much 
better quantity to analyze than the energy gap. In the next section we pursue this idea to 
investigate the phase diagram as a function of $\theta$.

\section{Numerical results}

\begin{figure}[htb]
\vskip .4cm
\includegraphics[scale=0.48]{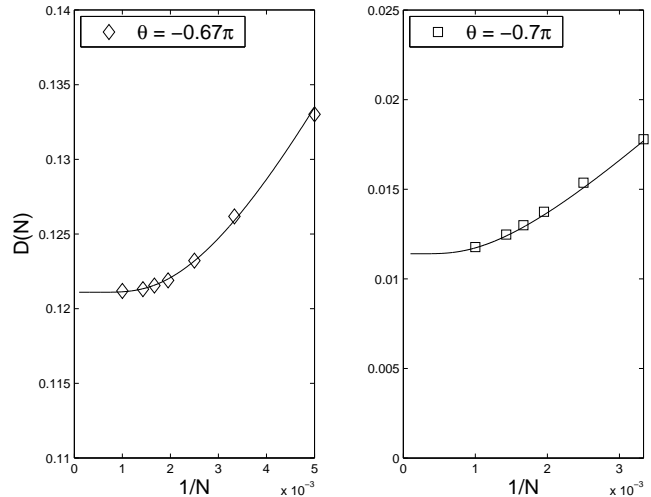}
\caption{Finite size scaling of the dimer-order parameter at $\theta = -0.67\pi$ and 
$-0.70\pi$ for $100\le N \le 1000$.}
\label{fig:dimer}
\end{figure}

For the reasons presented above, the highest chance to find the subtle nondimerized
spin-nematic phase near $\theta = -3\pi/4$ is by studying the dimer order in open chains.
Therefore, we calculated $D(N)$ using OBC up to 1000 lattice sites for various $\theta$
values between $-3\pi/4$ and $-\pi/2$. We have set $\chi\simeq 10^{-8}~$,
$M_{\rm min}=256$ and used three DMRG sweeps. Due to the very large correlation
length in the vicinity of the ferromagnetic phase boundary the maximum value of $M$ is
varied in the range of $600-1400$ for our longest chains. Our numerical results are shown
in Fig.~\ref{fig:dimer} for two different $\theta$ values. The numerical error is much
smaller than the size of the symbols.

\begin{figure}[htb]
\vskip .4cm
\includegraphics[scale=0.5]{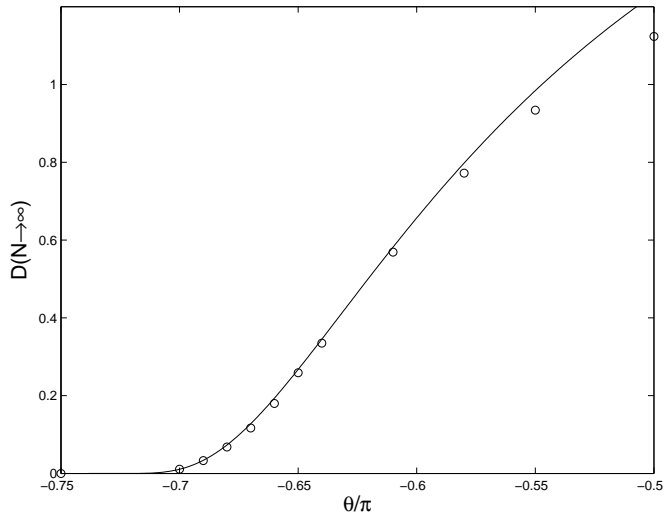}
\caption{The extrapolated value of the dimer order parameter as a function of
$\theta/\pi$. The solid curve is a least square fit using the form in Eq.\ (\protect\ref{eq:kt}) 
with $a=2.4\pm 0.6$, $c=0.11\pm 0.05$, and $\sigma=1.3\pm 0.3$.}
\label{fig:dimer_theta}
\end{figure}

It is seen in the figure that for large $N$ the data points show an upward curvature as
a function of $1/N$ for all $\theta$ values. On the other hand, the inflexion point
shifts towards very long chain lengths with decreasing $\theta$ values. The extrapolated
value $D$ of the dimer order parameter was determined using Eq.~(\ref{eq:scale_D}).
These values are shown in Fig.~\ref{fig:dimer_theta} as a function of $\theta$.
In fact, we found a finite, nonzero $D$ for all $\theta$ shown. However, we have no 
results for $\theta < 0.7\pi$, where $D(N)$ is so small that it is comparable to the 
numerical error. Although the dimer order parameter decreases very rapidly for 
$\theta<-0.64\pi$, the smooth behavior of $D$ as a function of $\theta$ and the upward 
curvature observed suggests that it vanishes at $\theta = -3\pi/4$.

It is also apparent from Fig.~\ref{fig:dimer_theta} that the dimer order parameter
resembles the form of the Berezinskii-Kosterlitz-Thouless (BKT) transition,
thus it opens exponentially slowly as a function of $\theta$,
\begin{equation}
D(\theta) = a \exp[-c(\theta-\theta_{\rm c})^{-\sigma}],\quad \theta_c=-3\pi/4,
\label{eq:kt}
\end{equation}
for $\theta>\theta_{\rm c}$. In Eq.~(\ref{eq:kt}), $a$ and $c$ are nonuniversal constants
and $\sigma$ is a characteristic exponent, which was estimated by a least square fit to 
be $\sigma=1.3\pm 0.3$ (see Fig.\ \ref{fig:dimer_theta}). This functional form is in 
qualitative agreement with our earlier numerical result.\cite{fath01}

Based on our calculations we conclude that there is no sign of any phase transition to
either a gapped or a gapless nondimerized phase in the vicinity of
$\theta\simeq-0.67\pi$, and thus a direct phase transition takes place
between the ferromagnetic and the dimerized phases.

\section{Conclusion}

In summary, we have performed a density-matrix renormalization-group calculation on the
spin-1 bilinear-biquadratic spin chain model in the vicinity of the ferromagnetic phase in 
order to search for a nondimerized quantum nematic phase suggested by 
Chubukov\cite{chubukov} or an extended critical region reported recently by L\"auchli 
{\sl et al.}\cite{lauchli} We took special care of the numerical accuracy since it has 
special importance in the present problem. We used the DBSS approach with the 
maximum number of block states varying between 1000 and 2000, and performed 
calculations on very long chains up to $N=1000$ lattice sites.

As a benchmark case we analyzed in detail the exactly solvable point $\theta=-\pi/2$.
We have found that the dimer order parameter is a quantity, which can be determined
numerically much more accurately than the energy gap, because of their unequal finite-size
scaling forms. Whereas the gap scales algebraically in open chains, the dimer order
parameter, which is a local quantity measured in the middle of the chain, becomes highly
independent of end effects and scales exponentially.

The phase diagram of the model was explored in general by computing the dimer order
parameter as a function of $\theta$. We have found strong indications that the dimer order,
which characterizes the dimer phase, only vanishes at the phase boundary of the
ferromagnetic phase $\theta = -3\pi/4$. The behavior  resembles that of
Berezinskii-Kosterlitz-Thouless transition, i.e., the order parameter opens exponentially
slowly as we move away from the transition point.  Our findings are in agreement with
the exponentially slow opening of the energy gap reported earlier.\cite{fath01}
Nevertheless, the actual transition itself is of first order as it involves direct level crossings,
as well. It is noteworthy that at $\theta = -3\pi/4$ the model has an extra SU(3)
symmetry,\cite{sutherland} which causes extra degeneracies in the spectrum. We are
tempted to speculate that this extra symmetry may play a role in that the two phases with
seemingly unrelated broken symmetries, SU(2) and translation invariance, meet in this
special point without an intermediate phase.

We could not find any trace of a nondimerized regime, at least surely not above
$\theta\approx 0.7\pi$. Below this limit numerical precision is a crucial issue since the
quantities of interests are extremely small. If the intermediate phase exists, it should be
constrained in a very narrow region near $\theta = -3\pi/4$, certainly much narrower
than predicted by L\"auchli {\sl et al.} The more likely alternative  interpretation, i.e.,
the nonexistence of the intermediate phase, is clearly at odds with current field theory
analysis. A reconciliation of the numerical results with field theory would be very welcome
in the future.

\acknowledgments

This research was supported in part by the Hungarian Research Fund(OTKA)
Grants No.\ T 043330, F 046356 and T 047003. The authors acknowledge
computational support from Dynaflex Ltd under Grant No. IgB-32.
\"O. L. was also supported by the J\'anos Bolyai scholarship.

\end{document}